\PassOptionsToPackage{dvipsnames}{xcolor}
\documentclass[sigconf, nonacm]{acmart}
\usepackage[ruled]{algorithm2e}
\usepackage{algorithmic}
\usepackage{multirow}
\usepackage{subcaption}
\usepackage{balance}
\usepackage{cleveref}
\usepackage{scalerel}
\newcommand{\smallast}{\scalebox{.7}{*}}
\newcommand\blfootnote[1]{%
  \begingroup
  \renewcommand\thefootnote{}\footnote{#1}%
  \addtocounter{footnote}{-1}%
  \endgroup
}
\usepackage{fancyvrb}

\newcommand\vldbpagestyle{plain} 

\begin{document}
\title[Zero-Execution Retrieval-Augmented Configuration Tuning of Spark Applications]{Zero-Execution Retrieval-Augmented Configuration Tuning \\of Spark Applications}

\author{Raunaq Suri,\smallast \quad Ilan Gofman,\smallast \quad Guangwei Yu,\quad Jesse C. Cresswell}
\affiliation{%
\institution{\scalebox{1.0}{Layer 6 AI, Toronto, Canada}}
\institution{\scalebox{1.0}{\string{raunaq, ilan, guang, jesse\string}@layer6.ai}}
}

\begin{abstract}
    
Large-scale data processing is increasingly done using distributed computing frameworks like Apache Spark, which have a considerable number of configurable parameters that affect runtime performance. For optimal performance, these parameters must be tuned to the specific job being run. Tuning commonly requires multiple executions to collect runtime information for updating parameters. This is infeasible for ad hoc queries that are run once or infrequently. Zero-execution tuning, where parameters are automatically set before a job's first run, can provide significant savings for all types of applications, but is more challenging since runtime information is not available. In this work, we propose a novel method for zero-execution tuning of Spark configurations based on retrieval. Our method achieves 93.3\% of the runtime improvement of state-of-the-art one-execution optimization, entirely avoiding the slow initial execution using default settings. The shift to zero-execution tuning results in a lower cumulative runtime over the first 140 runs, and provides the largest benefit for ad hoc and analytical queries which only need to be executed once. We release the largest and most comprehensive suite of Spark query datasets, optimal configurations, and runtime information, which will promote future development of zero-execution tuning methods.
    \blfootnote{* Equal Contribution}
\end{abstract}

\maketitle

\pagestyle{\vldbpagestyle}

\section{Introduction}
\label{sec:intro}
As demand for large-scale data processing continues to grow, there is an increasingly critical need for high-performance computing frameworks. Apache Spark \cite{SPARK/10.1145/2934664} and other distributed systems such as Hadoop \cite{hadoop} and Flink \cite{flink} have become central to managing large workloads. Each framework has its own set of configurable parameters that can have a large influence on runtime performance, but the optimal settings depend on the specifics of the job being run. Poor configuration parameter selection can result in under-utilization of system resources, or worse, over-commitment resulting in application failures \cite{cherrypick, pimpley2022towards}. Traditionally, optimization techniques involve manual parameter tuning by expert engineers, but this is complex, time-consuming, and error-prone.  Recent work has explored online tuning methods for Spark that require multiple executions of the same application to find optimal parameters \cite{locat, tencent, turbo, tuneful}. Information collected at runtime is used to propose a new set of parameters, and the job is rerun forming an interactive update loop. A common variant of online-tuning is one-execution tuning \cite{simtune}, where an application is run once using the default parameters, and performance metrics from that execution guide the selection of an improved configuration for the next run. Since these methods rely on metrics from the executions, all of them require at least one run of the job using default parameters before any tuning takes place, with the notion that upfront tuning costs can be amortized over future executions, leading to long-term savings. However, when multiple executions of the same query are not required, for example, when a data scientist runs an ad hoc query for dataset analysis, there is no benefit to online tuning. 

Parameter tuning still remains useful when queries only need to be run once or a few times, as long as tuning can be done before the first execution. The benefit is more evident for larger databases or computationally expensive queries, where a tuned query can be hours faster compared to the default settings. Zero-execution tuning is achievable by leveraging historical tuning data of similar queries and information about the job itself that is available pre-execution, such as the logical plan constructed by Spark SQL. Our research objective is to determine whether reframing parameter selection as a retrieval problem, rather than an optimization problem which is commonly done, can enable zero-execution tuning of Spark parameters. Towards this goal, we introduce \textbf{ZEST} (\textbf{Z}ero-\textbf{E}xecution \textbf{S}park \textbf{T}uning), a method that leverages historical data to retrieve the optimal configurations for similar queries prior to ever executing the target query. We demonstrate that ZEST achieves performance improvements over default parameters comparable to the gains from one-execution tuning and online optimization methods, but without access to any runtime information.

Our experimental evaluations across methods are done on the TPC-H \cite{tpch} and TPC-DS \cite{tpcds} datasets using the current version of Spark 3.5.3. We release our code for curating these datasets, along with the queries used to benchmark methods, and the optimal configurations needed for retrieval. In addition, we publish historical execution metrics of the queries to expedite future research in this space and to provide reproducible results at \\\href{https://github.com/layer6ai-labs/spark-retrieval-tuning}{\texttt{github.com/layer6ai-labs/spark-retrieval-tuning}}.

\section{Background}
\label{sec:background}
For the purposes of this work, we focus on a single data processing framework, Apache Spark, specifically Spark SQL \cite{sparkSQL}, given its widespread use in large-scale data analytics \cite{SPARK/10.1145/2934664}. Spark SQL operates by breaking down a query into multiple stages, each consisting of a set of tasks that are executed in parallel across a cluster of machines. When a query is submitted to Spark, it is first parsed into a \emph{logical plan}, which represents the high-level operations to be performed without specifying execution details. This logical plan is then optimized and transformed into a physical plan that outlines how operations will be executed across the cluster. The physical plan dictates how the query will be divided into jobs, stages, and tasks, which are then distributed to executors for parallel processing, ensuring efficient execution across the cluster. Our work shows how to leverage the logical plan, which is available prior to runtime and is hardware-agnostic, for zero-execution configuration tuning.

\subsection{Spark Configuration Tuning}

A Spark SQL application has over 200 tunable parameters that influence resource allocation, memory management, and execution strategies \cite{spark-conf-list}. Adjusting these settings for a given job can have a large effect on execution time and scalability. Hence, it is essential to tune these parameters based on the specific workload to be run for efficient resource utilization in distributed processing.

Previous work has focused on optimizing configuration parameters using online tuning methods \cite{tencent, locat, tuneful, turbo}.  These are centered on running variations of Bayesian optimization (BO) until a stop condition is met, which could be a predefined number of iterations, convergence criteria, or a computation budget. Given the substantial upfront costs associated with repeatedly executing the same job, it is not beneficial to apply BO for tuning jobs that will only be executed once or a limited number of times. Furthermore, various existing methods \cite{tencent, turbo} combine both offline and online tuning, but don't quantify the added benefit of the offline data, making it difficult to determine how much the offline component contributes to the tuning efficiency and overall performance gain. 

With the introduction of the Adaptive Query Execution (AQE) \cite{AQE} feature  in Spark 3.0, the effect of online tuning has been diminished. AQE enables runtime re-optimization which allows configuration parameters to be adjusted based on the operations outlined in the query execution plans.  This feature has reduced the need for manual intervention during the online tuning process as various configuration parameters are optimized before execution through native Spark features. Spark achieves this by applying a set of predefined rules at runtime, without relying on any historical data about the cluster or workloads.

Despite the clear benefits of using tuned parameter configurations from the very first run, to the best of our knowledge no recent research proposes zero-execution Spark tuning methods. The dominant approach, BO, inherently relies on the iterative process of rerunning the query to observe its runtime characteristics. Although our research objective is zero-execution tuning, we still compare our method against the performance of the BO-based optimization methods requiring at least one execution. We also compare to the heuristic-based zero-execution tuning method of using default parameters and relying on AQE.

\subsection{Embeddings for Tuning Methods}

Embedding-based retrieval methods have become widely used across domains due to their effectiveness in encoding complex datatypes such as text \cite{mikolov2013efficient} or code \cite{alon2018code2vec, NeelakantanCodeEmbeddings} into a dense vector representation. This allows for efficient similarity comparisons in the latent space, rather than relying on exact matching of inputs. Such methods can identify the closest vector based on Euclidean distance, but more commonly use similarity measures like cosine similarity which can be more computationally efficient when vectors are normalized. 

In the same way, embedding-based methods have proven to be effective for tuning distributed systems by mapping the query representations into a latent space. Query2Vec \cite{query2vec} demonstrated the feasibility of learning vector representations for SQL queries, and showed that the representations could be used for workload tasks and error prediction.  Qtune \cite{qtune} built off of Query2Vec to perform configuration tuning by converting SQL queries into vector representations using specific SQL features. It then employed the embeddings with reinforcement learning to learn the relationships between queries, database states, and configurations. For Spark, \citet{singh2022using} showed that learning embeddings and applying fuzzy matching can predict similarity between queries. However, for Spark configuration tuning specifically, previous methods have not used embedding-based approaches to represent the Spark SQL query, instead focusing on runtime features of the query which always require at least one execution to gather.

\section{Related Work}
\label{sec:related}

In this section, we discuss various approaches to Spark tuning present in the literature. Most Spark tuning methods are online and consist of two components: predicting a set of performant initial configuration parameters; and performing online tuning using an optimization method. To the best of our knowledge, no prior works use retrieval for Spark tuning.

\begin{figure*}[t]
    \centering
    \includegraphics[width=\linewidth, trim={5, 5, 5, 5}, clip]{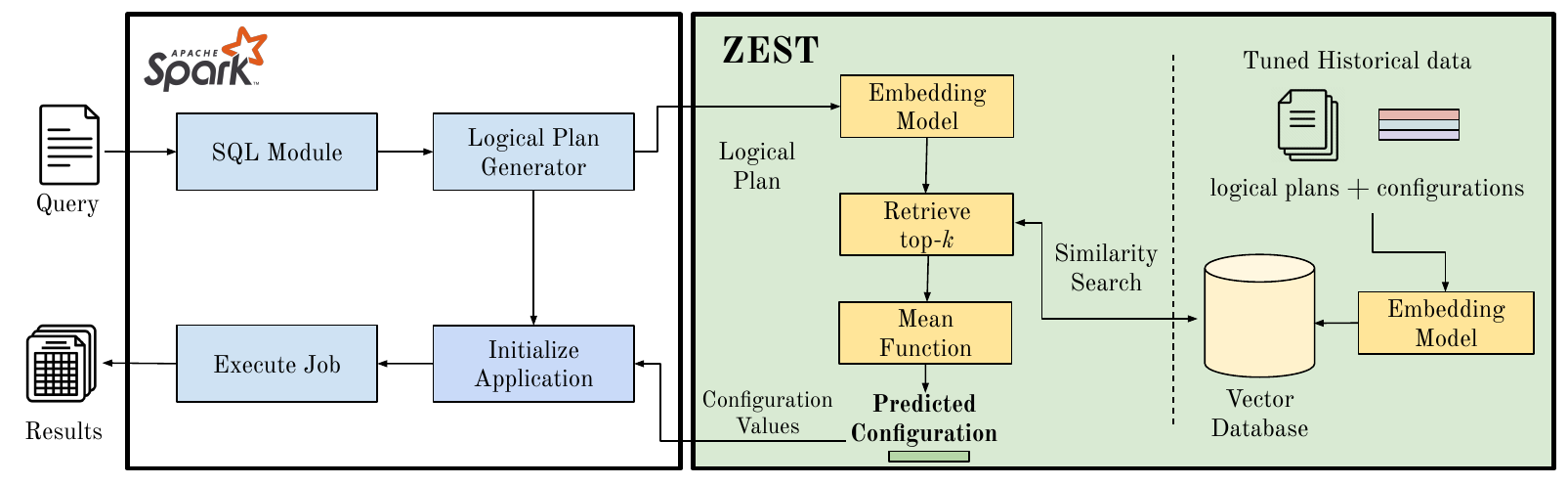}
    \caption{Overview of our zero-execution method for retrieval-augmented tuning of Spark configuration parameters. A query is first processed by the Spark SQL module to generate a logical plan. ZEST embeds the logical plan and retrieves the top-$k$ most similar historical embeddings with their corresponding tuned configurations from a vector database. ZEST returns the mean of retrieved configurations to Spark which then initializes a new application to execute the job.}
    \label{fig:ZEST-architecture}
    \Description{ZEST architecture for Spark Configuration Tuning. This figure presents the architecture of ZEST, a method designed for optimizing Spark Configuration parameters. The flow starts with a query that is processed through the Spark SQL module. The module generates a logical plan for the query which is passed into ZEST. ZEST embeds the logical plan using an embedding model. A similarity search is performed using the embedding and a vector database containing previously tuned historical logical plan and their optimal configurations. ZEST retrieves the top-k most similar queries and aggregates all similar queries values by taking a mean of the configuration parameters. These configuration parameter values are fed back into Spark, where they are used to initialize a new Spark application and execute the job with the optimized configuration.}
\end{figure*}

\subsection{Initial Configuration Prediction}\label{subsec:init_param_prediction}

The efficacy of online optimization greatly depends on the quality of the initial parameter set, so an explicit step to choose these parameters is usually necessary. Although existing methods refer to \textit{initial configuration} parameter selection, no method selects the parameters prior to the first execution of the Spark job. Instead, the starting point for online optimization is predicted with a model that uses runtime features \cite{yoro, simtune, tuneful, locat}. Each query in a training dataset is executed once using the default parameters of the Spark cluster. The runtime metrics of each query serve as features to train a model that predicts optimal starting configurations, and include key workload metrics such as input data size, shuffle operations, memory usage, and elapsed time. At inference time, any new query must also be executed once with default parameters before its features can be used to predict its optimal starting configuration. Crucially, the default run is likely far from optimized; indeed, we show that using default parameters can take more than 8$\times$ as long as optimal parameters for large data sizes.

\subsection{Online Parameter Tuning}

The majority of work on Spark tuning focuses on online tuning techniques, including BO, which refine the predicted initial configuration. These methods dynamically adjust configuration parameters across repeated executions to minimize subsequent runtimes. Much research has focused on improving the convergence speed of BO. Common approaches include pruning bad configurations before they are run \cite{turbo, tencent}, modifying the acquisition function \cite{locat}, and using an surrogate model in the place of actually executing Spark jobs \cite{yoro}. Additionally, some methods aim to improve convergence by choosing initial configurations close to the optimal, using either a global oracle model or measuring similarity between the target application and historical applications to guide configuration choices \cite{turbo, simtune, tuneful}.

Despite advancements in convergence speeds, online tuning remains expensive. The amortized cost of online tuning is high for the initial stages, since the configuration parameter space needs to be explored before convergence to the optimal values. For instance, the Tuneful method \cite{tuneful} requires at least 35 iterations before converging.  While more recent methods such as LOCAT \cite{locat} are more efficient, they still require at least 10 iterations. This makes online tuning not helpful for optimizing ad hoc queries, such as one-off analytical tasks, data exploration or exploratory queries done by data scientists, which are often run only once or a few times. 

Historical data of query executions can be leveraged to reduce the required number of iterations, for example by building a predictive model to simulate BO~\cite{yoro}. Other methods improve BO by using historical data to train a model that predicts similarity between configurations \cite{tencent}, or predicts if a configuration is suboptimal \cite{turbo}. One evident gap is the lack of transparency regarding the quantity and quality of the historical data being used. Based on published research it is often unclear how much historical data is required to produce reliable results, and how this historical data was obtained in the first place. Without explicit benchmarks that allow us to measure the amount of historical data used, it becomes difficult to extrapolate published results to new settings, and to compare to newly proposed methods. 

To the best of our knowledge all published Spark tuning methods use an online tuning component, apart from rules-based tuning \cite{ituned, self-tuning-fuzzy-rules}. This highlights the current dearth of effective techniques for zero-execution Spark tuning which can benefit ad hoc queries. Hence, in our experimental comparisons we compare our zero-execution method to one-execution methods that use only the initial configuration prediction component from recent works \cite{yoro, simtune, tencent}. Additionally, we compare these methods to BO, which, with sufficient iterations, can explore the configuration space and identify the optimal configuration, allowing us to benchmark against the best possible results.

\begin{table*}[t]
    \centering
              \setlength{\tabcolsep}{2pt}
    \caption{Spark Configuration Search Space}
    \begin{tabular}{llccc}
        \toprule
        Spark Configuration Parameter & Description & Default Value & EMR Range & Local Cluster Range \\
        \midrule
        spark.sql.shuffle.partitions & \small{Number of partitions for shuffle operations} & 200 & [50, 1000] & [50, 1000]  \\
         spark.executor.instances & \small{Maximum number of executors} & 4 & [1, 28] & [1, 180] \\
         spark.driver.memory & \small{Amount of memory (GB) allocated to the driver process} & 1 & [1, 44] & [1, 9] \\
         spark.driver.cores & \small{Number of CPU cores allocated to the driver process} & 1 & [1, 7] & [1, 5] \\
         spark.executor.memory & \small{Amount of memory (GB) allocated to each executor process} & 1 & [1, 44] & [1, 60] \\
         spark.executor.cores & \small{Number of CPU cores allocated to each executor process} & 1 & [1, 7] & [1, 10] \\  
        \bottomrule
    \end{tabular}
    \label{tab:search-space}
\end{table*}

\section{Zero-Execution Configuration Tuning Using Retrieval}
\label{sec:method}

Embedding-based retrieval methods have been successfully used in a variety of domains such as question answering \cite{embedding-based-retrieval}, recommendation systems \cite{zhao2023embeddingrecommendersystemssurvey}, and semantic search \cite{muennighoff2022sgptgptsentenceembeddings}, demonstrating their versatility in retrieving relevant information.  We propose embedding logical plans generated by Spark and applying retrieval to predict configuration parameters before ever running the Spark job. Logical plans are standardized and outline the exact set of internal operations to be executed within the application, making them a rich source of information entirely available before runtime. Plans with related operations and structure are likely to exhibit comparable behavior, hence, we propose that their optimal parameter configurations would also likely align. In this section, we describe how to create a diverse retrieval index of optimal configuration settings from historical data, similar to how existing optimization methods use historical data to train predictive models (\Cref{subsec:init_param_prediction}). Then we show that semantic embeddings based on logical plans are effective for retrieval-augmented tuning. An overview of our proposed zero-execution Spark tuning method, ZEST, is shown in ~\Cref{fig:ZEST-architecture}.

\subsection{Logical Plan Execution Index Creation}\label{index-creation}
To evaluate zero-execution tuning, we concentrate on the types of analytics workflows that are frequently run on large enterprise databases to support business decision making. To generate an index of optimal configurations that represents the myriad of analytics and data processing queries that would be run in a large organization, we use queries from the TPC-H \cite{tpch, tpch-spark} and TPC-DS \cite{tpcds, tpcds-spark} benchmarks. TPC-H is composed of 22 queries representing business-oriented ad hoc queries that have industry-wide relevance \cite{tpch}. TPC-DS contains a set of 99 queries which encompass representations of reporting queries, ad hoc queries, online analytical processing queries, and data mining queries. From these query datasets, we generate logical plans and find optimal parameter configurations, the latter of which can also be used to evaluate the performance of all other methods we compare to.

Parameter tuning is an exponentially hard problem given the multiplicative nature of possible combinations as the number of tuned parameters grows. Though there are hundreds of Spark configuration parameters which can affect the execution behaviour of a query \cite{spark-conf-list}, prior research has shown that only a small subset of the parameters have a significant impact on the runtime of a job \cite{locat, tencent, tuneful}. \Cref{tab:search-space} shows the six parameters composing the search space that we investigate, and the ranges of parameters we search over for two cluster configurations -- a standardized Amazon Elastic MapReduce (EMR) cluster and a local compute cluster available to us. For the initial data collection of optimal parameter configurations, we used BO based on Tree-structured Parzen Estimators as implemented in Optuna \cite{optuna}. Each query was optimized for four different input data sizes (100GB, 250GB, 500GB, and 750GB), starting from the default Spark parameters and running for 40 iterations of BO to minimize runtime.

We developed a listener that would capture the logical plan of each application, as well as execution metrics for each iteration of BO upon successful execution \cite{spark-measure, yoro}. After capturing data for all 121 queries in TPC-H and TPC-DS, 4 data sizes, and 40 iterations per query, our resultant dataset consisted of 19,360 parameter configurations, logical plans, and their corresponding execution metrics. We have released this dataset to facilitate future research involving Spark applications. The dataset was partitioned into a 90/10 train/test split by query in order to prevent test data leakage; the same splits were used to generate our retrieval index and for the purposes of training initial configuration prediction methods as a baseline (\Cref{subsec:init_param_prediction}).

\begin{figure*}
    \centering
    \includegraphics[width=\linewidth, trim={6, 8, 9, 6}, clip]{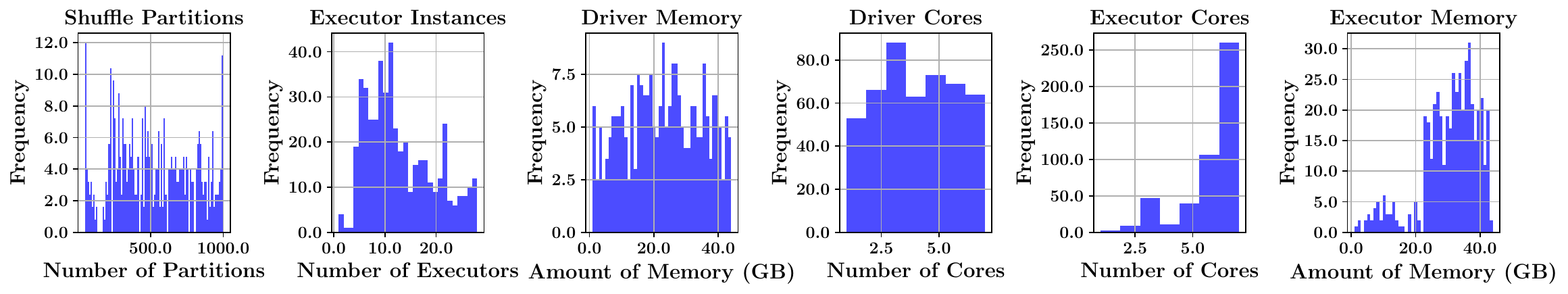}
    \caption{Distribution of configuration values for the optimal parameters for a query in the index on the EMR cluster. The x-axis denotes the values of the configuration parameters and the y-axis denotes the frequency of those parameter values in the optimal configuration across all queries and input data sizes.}
    \label{fig:best-params-distribution}
    \Description{A collection of 6 histograms, each corresponding to a parameter from {\Cref{tab:search-space}}. The histograms show that there is variance in the values for each parameter across the index.}
\end{figure*} 

For each query-size pair in the training set, we generated an embedding for its logical plan via a text embedding model. The retrieval index was then populated with a mapping of embeddings to their corresponding configurations which yielded the lowest runtime. \Cref{fig:best-params-distribution} shows the distribution of the optimal parameter values for each (query, size) pair used to populate the index. From the histograms we see that there is significant variance in the optimal value of each parameter. This demonstrates the need for per-query tuning, as no single configuration is optimal for all queries. The histograms also show that our dataset of optimized query executions covers a wide range of configuration parameter values. We also observed different distributions across parameters -- the optimal number of executor cores tends to be the maximum possible value for most queries, but the optimal executor memory has a bimodal distribution. Certain configuration values such as the number of shuffle partitions and the driver memory do not have a clearly discernible pattern that is easily captured by heuristics.

\subsection{Optimizing Configurations using Retrieval}
Given a new query, we use the same embedding model to convert its logical plan to a vector in the embedding space. We then compute cosine similarities with the indexed vectors, and retrieve the top-$k$ most similar queries, along with their corresponding optimal configurations. To predict a configuration for the new query, we calculate the parameter-wise mean of the $k$ retrieved configurations, as shown in \Cref{alg:knn}. Notably, since our retrieval is based on logical plans which are available before runtime, ZEST generates optimized configurations without executing the new query, in contrast to all state-of-the-art methods which use at least one execution (\Cref{sec:related}). While ZEST does rely on an index of historical query executions, this is equivalent to how previous works train a predictive model on historical executions (\Cref{subsec:init_param_prediction}).

\begin{algorithm}[t]
    \caption{Zero-Execution Spark Tuning using Embedding-based Retrieval}
    \label{alg:knn}
    \begin{algorithmic}[1]
        \REQUIRE Query logical plan $\ell$, Index $I$, Embedding model $E$, Neighbour count $k$.
        \ENSURE Predicted configuration $\hat{C}$.
        
        \STATE \textbf{Embedding Generation:} Generate embedding $\hat e \gets E(\ell)$,\\  a vector representation of the query logical plan.
        \STATE \textbf{Similarity Search:} For each embedding $e_i \in I$, compute \\ the cosine similarity $s_i = \frac{\hat e \cdot e_i}{\|\hat e\|\|e_i\|}$.
        \STATE \textbf{Configuration Recommendation:} Return as $\hat{C}$ the parameter-wise mean of configurations associated with the top $k$ embeddings by $s_i$.
    \end{algorithmic}
\end{algorithm}

Our method relies on two important hyperparameters to compute the recommended configuration: the embedding model, and the value of $k$. Both of these directly impact the performance of the method when applied to the pre-generated index. To maximize the effectiveness of the embedding retrieval, it is important to choose a model that captures the most relevant semantics and structures of Spark logical plans. Though there are no open-source embedding models trained specifically for Spark logical plans, there are many available models that can be used for general text and coding tasks. From the top performing text embedding models on the MTEB leaderboard \cite{mteb}, we tested five options: \texttt{jina-embeddings-v3} \cite{jina-embeddings}, \texttt{stella\_en\_400M\_v5} \cite{stella_en_400M_v5}, \texttt{UAE-Large-V1} \cite{whereisai-uae-large-v1}, \texttt{bge-large-en-v1.5} \cite{bge_embedding}, and  \texttt{gte-large-en-v1.5} \cite{gte1, gte2}. We performed hyperparameter tuning with 10-fold cross-validation and found $k=29$ with the \texttt{jina-embeddings-v3} model to perform the best. We omit detailed results for brevity, but release the code to reproduce these steps.

\section{Experiment Setup}
\label{sec:experiment-setup}

The Amazon EMR cluster we used was composed of one main \texttt{r6g.2xlarge} node and four \texttt{r6g.2xlarge} core nodes, each running EMR version 7.1.0, resulting in a server composed of 40 CPU cores and 300 GB of memory \cite{r6g}.  To check for the robustness of the methods presented, the index was also recreated for a parallel set of experiments in a local cluster setup within an internal network. For comparison, the local cluster was composed of 4 machines running in a Docker Swarm. Each machine had two worker nodes resulting in 8 total workers with each node having 70 GB of memory and 15 cores. One machine also served as the driver node using 10 GB of memory and 5 cores. This resulted in a local cluster setup composed of 570GB of memory and 125 CPU cores. We repeat all the experiments in the standardized cloud-based environment, and our local environment. 

We compare ZEST to the initial configuration prediction methods reviewed in \Cref{sec:related}, namely SimTune \cite{simtune}, YORO \cite{yoro}, and the method by \citet{tencent}, all of which require one execution of the query using default parameters to generate runtime features. Some of the baseline methods have not open-sourced their code, models, or training data~\cite{yoro, tencent}, so we have implemented them to the best of our abilities based on their published descriptions, and trained predictive models using our train/test split. We also evaluate against the obtained optimal configurations using BO with Optuna over 40 iterations, the same method that originally produced these configurations for our index. Lastly, we also compare with a heuristic zero-execution method of simply using the default Spark parameters \cite{spark-conf-list} (shown in \Cref{tab:search-space}) and relying on AQE for optimization.

For our performance comparisons we evaluate on the test split of 10\% of queries from the TPC-DS and TPC-H datasets, across the four input data sizes defined in \Cref{index-creation}. ZEST and all baseline methods, except for Optuna, rely on a historical dataset of executed queries for which we used the 90\% of queries in the training split. For each test (query, size) pair, we ran each method 10 times and averaged the results to mitigate randomness in the computing environment.

\section{Analysis}
\label{sec:analysis}

\subsection{Evaluation Against Baselines}\label{sec:overall_timing}

\subsubsection{Overall Performance}\label{subsec:overall_perf}

The test queries were evaluated against both the EMR setup and the local cluster setup as specified in \Cref{sec:experiment-setup}. The results of all test (query, size) pairs were then summed to determine the overall runtime performance of the method. In all cases, lower runtimes are desirable. The runtime performance on the EMR cluster is shown in \Cref{tab:overall-results}, while the local cluster is shown in \Cref{tab:overall-results-cluster}. We bold the best results out of each group of methods that uses the same number of test query executions $N$ prior to the final execution which runtime is measured for.

\begin{table}[t]
    \centering
    \caption{Total execution time in seconds on TPC-DS and TPC-H datasets using the EMR cluster. $N$ denotes the number of executions of the test query used prior to the final result.}
    \setlength{\tabcolsep}{10pt}
    \begin{tabular}{lrrrr}
        \toprule
        \textbf{Method} & $N$ & TPC-DS &  TPC-H & Total 
         \\
         \midrule
         Optuna \cite{optuna} & 40 &  \textbf{956}  & \textbf{1,255} & \textbf{2,211} \\
        \midrule
        SimTune \cite{simtune} &1 & \textbf{966} & \textbf{1,288} & \textbf{2,254} \\
        YORO \cite{yoro} &1 & 1,909 & 1,579 & 3,488\\
        \citet{tencent} &1 & 970 & 1,805 & 2,775 \\
        \midrule
        Default \cite{spark-conf-list} & 0  & 7,008 & 9,953 & 16,961  \\
        \textbf{ZEST (Ours)} & 0 & \textbf{1,005} & \textbf{1,364} & \textbf{2,369} \\
        \bottomrule
    \end{tabular}
    \label{tab:overall-results}
\end{table}
\begin{table}[t]
    \centering
    \caption{Total execution time in seconds on TPC-DS and TPC-H datasets using a local cluster. $N$ denotes the number of executions of the test query used prior to the final result.}
    \setlength{\tabcolsep}{10pt}
    \begin{tabular}{lrrrr}
         \toprule
        \textbf{Method} & $N$ & TPC-DS &  TPC-H & Total \\
        \midrule
         Optuna \cite{optuna} & 40 &  \textbf{4,124} & \textbf{8,169} & \textbf{12,293}\\
         \midrule
        SimTune \cite{simtune} & 1 & \textbf{4,782} & \textbf{8,255} & \textbf{13,037}\\
        YORO \cite{yoro} & 1  & 6,562  &  10,644 & 17,206\\
        \citet{tencent} & 1 & 5,705 & 9,497 & 15,202\\
        \midrule
        Default \cite{spark-conf-list} &0  & 7,763 & 15,097 & 22,860 \\
        \textbf{ZEST (Ours)} &0 & \textbf{4,604} & \textbf{8,968} & \textbf{13,572}\\
        \bottomrule
    \end{tabular}
    \label{tab:overall-results-cluster}
\end{table}

From \Cref{tab:overall-results} we see that BO with 40 query executions results in the most optimized parameters, but of course this amount of optimization is highly impractical. In comparison, the default parameters are far from optimal, leading to runtimes about 7$\times$ as long. This point is crucial however, as all tuning methods other than ZEST run the test queries at least once with the default parameters to collect runtime information. Out of the one-execution methods, SimTune \cite{simtune} most closely approaches the lower bound set by BO. By leveraging similarities between logical plans which are available pre-execution, ZEST approaches the performance of highly optimized queries, coming within 8\% of extensive BO. We can also see that our method performs slightly better on the analytics and ad hoc queries present in TPC-H (speedup of 7.3$\times$ over default) than the broader query set found in TPC-DS (speedup of 7.0$\times$). This is promising, as analytics and ad hoc queries are the most likely queries to be run as one-offs, where zero-execution methods are the only sensible tuning options.

In the local cluster setting (\Cref{tab:overall-results-cluster}), ZEST continues to perform well compared to other methods. While the total runtimes differ between the EMR and local cluster, the relative ranking of methods remains consistent. ZEST remains competitive to the best one-execution approach, which reinforces the effectiveness of our configuration retrieval strategy across cluster setups. 

\subsubsection{Performance by Input Data Size}\label{subsec:perf_by_size}
Outside of the variance in query complexity and query types, the input data size can also factor into the performance of tuning algorithms. The nonzero-execution methods rely on the input data size as a key feature, which is information that must be collected during the first execution \cite{simtune, locat, tencent, turbo, yoro}. \Cref{tab:overall-results-by-size} shows the prediction performance of the methods across input data sizes. We see that the greatest performance improvements relative to the default settings are found for the largest input data size. This is likely because larger input sizes are more sensitive to the Spark configuration as they are at higher risk of encountering memory or CPU bottlenecks. Despite ZEST not having access to input data size as a feature, we do not observe a strong effect of size on its relative performance to Optuna or SimTune. This could indicate that input data size is no longer a significant factor in configuration tuning with the advent of Adaptive Query Execution built into Spark 3+ \cite{AQE}. AQE allows for dynamic shuffling of partitions, which is a significant bottleneck in processing larger data sizes, as miscalculated partition sizes can result in frequent spills of data to disk if the partition sizes are too large, and increased network I/O if partition sizes are too small. 

In summary, by reframing configuration tuning as a retrieval problem instead of an optimization problem, our method using similarity of logical plans achieves performance close to one-execution and multi-execution tuning without prior executions of the test query.

\begin{table}[t]
    \centering
    \caption{Performance of tuning methods by input data size on the EMR cluster, measured by total query execution time in seconds. $N$ denotes the number of executions of the test query used prior to the final result.}
    \begin{tabular}{lrrrrr}
        \toprule
        \textbf{Method} & $N$ & 100GB & 250GB & 500GB & 750GB \\
        \midrule
         Optuna \cite{optuna} & 40 & \textbf{209} & \textbf{373} & \textbf{678} & \textbf{950} \\ \midrule
        SimTune \cite{simtune} & 1 & \textbf{211} & \textbf{365} & \textbf{743} & \textbf{936} \\
        YORO \cite{yoro}& 1 & 290 & 637 & 1,023 & 1,538 \\
        \citet{tencent} & 1 & 219 & 475 & 853 & 1,228 \\
        \midrule
        Default  \cite{spark-conf-list} & 0  & 915 & 2,410 & 5,115 & 8,520 \\
        \textbf{ZEST (Ours)} & 0 & \textbf{205} & \textbf{386} & \textbf{711} & \textbf{1,066}\\
        \bottomrule
    \end{tabular}
    \label{tab:overall-results-by-size}
\end{table}

\subsection{Accumulated Cost of Query Executions}
In certain settings, such as production environments, multiple executions of a query over time are required. The appropriate metric for the performance of a tuning methodology is not the runtime of its most optimal configuration (\Cref{sec:overall_timing}), but the total cluster execution time for all tuning steps, and all subsequent production runs with the optimized configuration. We refer to this as the accumulated cost of query executions. Although Optuna and SimTune have slightly better performance than ZEST in terms of optimized configuration runtime (\Cref{tab:overall-results,tab:overall-results-cluster}), they only achieve this level after executing the query at least once with the default configuration. This can be expensive, as we saw the runtime with default settings can be over 7$\times$ as long as the optimized runtime (\Cref{tab:overall-results}). 
Hence, there is an up front cost of the initial query execution that must be amortized over a large number of production iterations in order to reap any benefits from tuning \cite{simtune}.

\begin{figure}[t]
    \centering
    \includegraphics[scale=0.75]{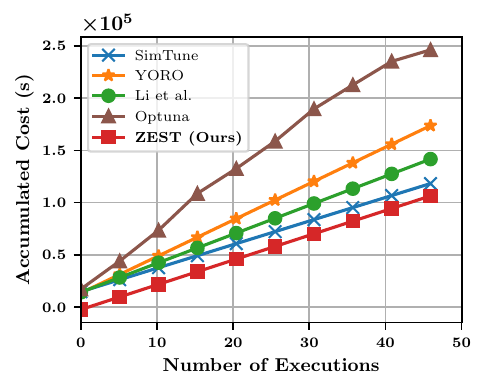}
    \caption{Accumulated cost of query executions on the TPC-H and TPC-DS benchmarks under different tuning algorithms on the EMR cluster. The x-axis does not count the initial run with default parameters used by methods other than ZEST.}
    \label{fig:amortized-cost}
\end{figure}

\Cref{fig:amortized-cost} shows the accumulated cost of query executions for the methods in \Cref{tab:overall-results} on the EMR cluster. For zero- and one-execution methods, the $x$-axis is the number of repeated executions of all (query, size) pairs in the test set using the tuned configurations, while for the Optuna method the configuration is updated every execution. The accumulated cost on the $y$-axis includes the cost of tuning (i.e. the runtime of the default configuration for one-execution methods), and subsequent query executions with the tuned configuration parameters. We see that ZEST remains the most cost effective solution after 50 repeated executions even though other methods eventually find more optimized configurations because ZEST's up front cost is negligible. The left-hand region of the plot around one single execution emphasizes that ZEST is by far the best choice for ad hoc queries.

\Cref{tab:amortized-intersection} shows the number of repeated executions needed for other methods to become more cost effective than ZEST. Only Simtune and Optuna find more optimized configurations; all other methods remain inferior to ZEST regardless of the number of executions. Still, the initial cost of executing with default configurations is so high that Spark queries which will be run fewer than 148 times would benefit more from our zero-execution tuning method than any one-execution method.

\begin{table}
    \centering
    \caption{Number of repeat executions required to be more cost effective than ZEST on the EMR cluster. $N$ denotes the number of executions of the test query used prior to the final result.}
    \setlength{\tabcolsep}{10pt}
    \begin{tabular}{lcc}
        \toprule
        \textbf{Method} & $N$ & Repeated Executions \\
        \midrule
         Optuna \cite{optuna} & 40 & 911 \\
        \midrule
        SimTune \cite{simtune} & 1 & 148 \\
        YORO \cite{yoro} & 1 & Always inferior \\
        \citet{tencent} & 1 & Always inferior \\
        \midrule
        Default  \cite{spark-conf-list} & 0  & Always inferior \\
        \bottomrule
    \end{tabular}
    \label{tab:amortized-intersection}
\end{table}

\subsection{Effect of Index Quality on Performance}
Methods such as SimTune \cite{simtune}, YORO \cite{yoro}, that of \citet{tencent} center around training predictive models. Such methods are sensitive to the size and quality of the historical data available for training. In an equivalent manner, ZEST uses historical data to create an index which underlies its performance at inference time. For retrieval-based methods, lower index quality can directly cause a drop in performance due to not having sufficient relevant items present. This section highlights ablations on the index to evaluate the methods' sensitivity to dataset quality.

\subsubsection{Generalization to new dataset sizes}
In the literature, it is often mentioned that, compared to offline tuning, online tuning has advantages as methods like BO are able to adapt to changing data sizes \cite{locat, tencent, tuneful, turbo}. To validate zero- and one-execution methods in settings with dynamic input data sizes, we reindex or retrain the methods on a training set composed only of queries with input data sizes of 100GB and 250GB, and test their predicted configurations against queries with input data sizes of 750GB. \Cref{tab:unseen-data-size} shows the results when queries with 750GB input data sizes are or are not in the historical data. We can see that most methods can robustly tune queries with much larger data sizes than what appears in their historical data. The results are largely similar, further supporting our conclusion that input data size is no longer a significant factor in the tuning process with the advent of AQE \cite{AQE}. Zero-execution tuning remains suitable, even with unexpectedly large input data sizes.

\begin{table}
    \setlength{\tabcolsep}{10pt}
    \centering
    \caption{Performance of tuning methods on queries with previously seen and unseen database sizes on EMR, measured by total query runtimes (seconds). $N$ denotes the number of executions of the test query used prior to the final result.}
    \begin{tabular}{lccc}
        \toprule
        \textbf{Method} & $N$ & 750 GB Seen & 750 GB Unseen \\
        \midrule
         Optuna \cite{optuna} & 40 & N/A & \textbf{950} \\
        \midrule
        
        SimTune \cite{simtune} & 1 & 936 & 957 \\
        YORO \cite{yoro} & 1 & 1538 & 1684 \\
        \citet{tencent} & 1 & 1228 & \textbf{933}\\
        \midrule
        Default \cite{spark-conf-list}  & 0  & 8520 & 8520 \\
        \textbf{ZEST (Ours)}  & 0 & 1066 & \textbf{1060} \\
        \bottomrule
    \end{tabular}
    \label{tab:unseen-data-size}
\end{table}

\subsubsection{Generalization to an Unseen Data Catalog}
The historical datasets used to populate the index or train configuration prediction models were sourced from two data catalogs: TPC-DS \cite{tpcds} and TPC-H \cite{tpch}. Each of those catalogs has its own set of tables with their respective schemas. To simulate applying tuning algorithms on unseen databases, we retrained or reindexed the methods with only queries from TPC-DS, and tested against TPC-H using the data collected from the EMR cluster.

\begin{figure*}[t]
    \centering
    \begin{subfigure}[t]{0.45\textwidth}  
        \includegraphics[scale=0.75]{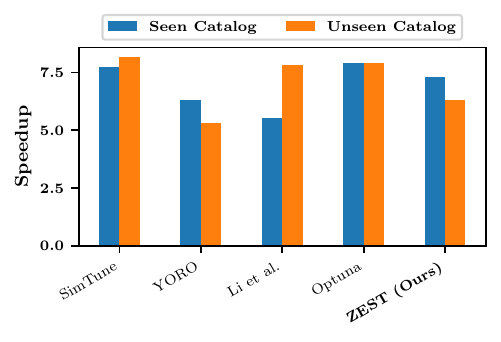}
        \caption{Tuning speedup compared to default parameters.}
        \label{fig:dataset-ablate-speedup}
    \end{subfigure}
    \hfill      \begin{subfigure}[t]{0.45\textwidth} 
        \includegraphics[scale=0.75]{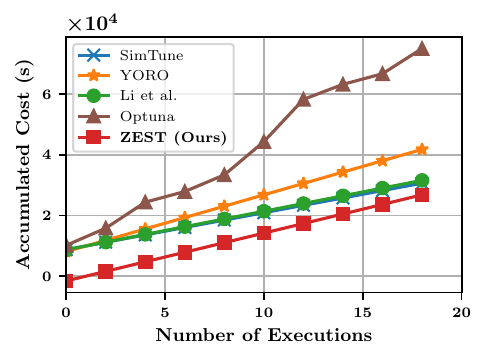}
        \caption{Accumulated cost of TPC-H query executions.}
        \label{fig:dataset-ablate-amortized}
    \end{subfigure}
    \caption{Tuning queries on an unseen data catalog. The methods were built on solely the TPC-DS data catalog and queries, and tested on the TPC-H data catalog and queries on the EMR cluster.}
    \label{fig:dataset-ablation}
\end{figure*}

\Cref{fig:dataset-ablate-speedup} shows the speedup of the TPC-H queries, which is computed as the ratio of the query runtime using default parameters, to the runtime after tuning. We see that ZEST experiences a minor drop in performance on the unseen catalog compared to results where the index contains queries from both data catalogs. This is expected because ZEST makes use of logical plans which contain column and table names, boosting the similarity score for queries executed on the same tables. For an unseen catalog, the text embedding models can only rely on the operations being executed to cluster similar queries. Executing against an unseen data catalog does not affect other methods as much, since those methods are largely based on execution metrics, which themselves are less dependent on the catalog specifics. SimTune \cite{simtune} and \citet{tencent} may have seen an improvement because TPC-H is a less noisy evaluation set.

However, the accumulated cost still favours ZEST for queries that are not repeated extensively, as seen in \Cref{fig:dataset-ablate-amortized}. The accumulated cost is still lower than any one-execution method for the first 30 repeated executions, making ZEST a better choice for ad hoc analytics queries, but not necessarily the most suitable for heavily repeated executions on an unseen catalog. The benefit of ZEST is most apparent for queries executed on databases that have previously been seen (\Cref{fig:dataset-ablate-speedup}, ``Seen Catalog''). In these cases, the retrieval components can leverage both the logical data operations and the execution history of previous queries on the same tables. This scenario is common in enterprises, where the same databases are frequently used for large-scale analytical tasks. With relevant historical data, ZEST becomes an even more effective choice in real-world settings where existing databases are repeatedly queried for new workloads.

\newcommand{\thincolorbox}[2]{%
  \colorbox{#1}{\raisebox{-0.2pt}[0.2\baselineskip][0\baselineskip]{#2}}%
}

\begin{figure*}[t]
    \centering
    \Description{Two SQL query templates are shown, one for the test set query TPC-H Query 21 and another for its closest query found by the embedding model, TPC-H Query 2. The queries look slightly similar, but have vastly different optimal runtimes.}
    \begin{subfigure}[t]{0.47\textwidth}
    {\footnotesize
    \begin{Verbatim}[commandchars=\\\{\}]
+----------------------------------------------------------+
| Default: 2652s, Optuna: 340s, SimTune: 389s , ZEST: 353s |
+----------------------------------------------------------+ 
\thincolorbox{SeaGreen}{Select} \textcolor{Emerald}{s_name}, count(*) as numwait from \textcolor{ForestGreen}{supplier}, lineitem l1,
orders, \textcolor{ForestGreen}{nation} \thincolorbox{YellowOrange}{Where} \textcolor{Mahogany}{s_suppkey = l1.l_suppkey} 
and o_orderkey = l1.l_orderkey and o_orderstatus = 'F'
and l1.l_receiptdate > l1.l_commitdate
and exists ( \thincolorbox{GreenYellow}{Select} * from lineitem l2 \thincolorbox{YellowOrange}{Where}
l2.l_orderkey = l1.l_orderkey and l2.l_suppkey <> l1.l_suppkey
) and not exists (\thincolorbox{GreenYellow}{Select} * from lineitem l3 \thincolorbox{YellowOrange}{Where}
l3.l_orderkey = l1.l_orderkey and l3.l_suppkey <> l1.l_suppkey
and l3.l_receiptdate > l3.l_commitdate)
and \textcolor{Mahogany}{s_nationkey = n_nationkey} and n_name = '[NATION]'
group by s_name \thincolorbox{Lavender}{order by} numwait desc,\textcolor{Fuchsia}{s_name};
    \end{Verbatim}
    }
    \vspace{1.3em}
    \subcaption{TPC-H Query 21, 500GB.}
    \end{subfigure}
    \hfill
    \begin{subfigure}[t]{0.47\textwidth}
    {\footnotesize
    \begin{Verbatim}[commandchars=\\\{\}]
+------------------------------------------------+
| Default: 220s, Optuna: 29s SimTune: -, ZEST: - |
+------------------------------------------------+ 
\thincolorbox{SeaGreen}{Select} s_acctbal, \textcolor{Emerald}{s_name}, n_name, p_partkey, p_mfgr,
s_address, s_phone, s_comment
from part, \textcolor{ForestGreen}{supplier}, partsupp, \textcolor{ForestGreen}{nation}, region
\thincolorbox{YellowOrange}{Where} p_partkey = ps_partkey and \textcolor{Mahogany}{s_suppkey = ps_suppkey}
and p_size = [SIZE] and p_type like '%[TYPE]'
and s_nationkey = n_nationkey and n_regionkey = r_regionkey
and r_name = '[REGION]'
and ps_supplycost = (\thincolorbox{GreenYellow}{Select} min(ps_supplycost) from partsupp, 
\textcolor{ForestGreen}{supplier}, \textcolor{ForestGreen}{nation}, region 
\thincolorbox{YellowOrange}{Where} p_partkey = ps_partkey and \textcolor{Mahogany}{s_suppkey = ps_suppkey}
and \textcolor{Mahogany}{s_nationkey = n_nationkey}
and n_regionkey = r_regionkey and r_name = '[REGION]')
\thincolorbox{Lavender}{order by} s_acctbal desc,n_name,\textcolor{Fuchsia}{s_name},p_partkey;
    \end{Verbatim}
    }
    \vspace{-0.56em}
    \subcaption{TPC-H Query 2, 500GB.}
    \end{subfigure}
    \caption{Template for TPC-H Query 21 from the test set and the corresponding most similar item in the retrieval index, TPC-H Query 2. Highlighted text indicates common query operations, while coloured text indicates the common columns on which they were executed.}
    \label{fig:qualitative-result}
\end{figure*}

\subsection{Qualitative Similarity}
ZEST has comparable performance to existing one- and multi-execution methods, providing evidence that the embedding models can identify unique characteristics in Spark logical plans to effectively gauge similarity between queries. Figure \ref{fig:qualitative-result} shows a sample test query and the corresponding most similar query from the retrieval index. We see that the query selected by the embedding model is not exactly similar in text, but is similar in operation. TPC-H Query 21 identifies suppliers who were not able to ship parts in a timely manner, while TPC-H Query 2 retrieved from the index finds a supplier to place an order for a given part in a given region \cite{tpch}. Both queries act on a similar set of tables and apply joins based on the \texttt{supplier} key and the \texttt{nation} key from the \texttt{supplier} table, with differences between the queries present in the aggregations, additional filtering, and the specific nested conditions. The queries have different runtimes, but their underlying similarity meant that they were sensitive to the same configuration parameters. As a result, their tuned configurations were closely aligned.

\section{Conclusion}
\label{sec:conclusion}

This paper proposes ZEST, zero-execution Spark tuning, which is retrieval-based and compares similarities through text embeddings of Spark logical plans. Without prior executions of a test query, ZEST is able to achieve 93.3\% of the performance improvements gained from one-execution and multi-execution tuning algorithms. This is especially advantageous for ad hoc and analytical queries which are not repeated frequently, resulting in significantly lower accumulated costs of query execution. We analyzed the robustness of ZEST to varying input data sizes and unseen databases, finding in each case significant benefits compared to the default parameters, and to one-execution methods in terms of accumulated cost.

As part of this work, we compiled a dataset of 19,360 query executions based on TPC-H and TPC-DS, including highly optimized parameter configurations and execution metrics. While we used this dataset for a retrieval index based on logical plans, it is equally applicable as a training dataset for predictive models, which are a common approach to Spark tuning. Hence, in the spirit of openness and reproducibility, we are releasing the dataset to facilitate further research in Spark tuning.

\begin{acks}
    The authors would like to thank Laura Doktorova for her assistance in the implementation of one of the baselines.
\end{acks}

\bibliographystyle{ACM-Reference-Format}
\bibliography{bib}

\end{document}